\documentclass[aps,pra,twocolumn,showpacs,superscriptaddress]{revtex4}

\usepackage{graphicx}
\usepackage{amsmath}
\usepackage{amssymb}

\input{epsf}
\begin{document}

\title{BEC-BCS Crossover of a Trapped Two-Component Fermi Gas
with Unequal Masses}

\author{J. von Stecher}
\affiliation{JILA and Department of Physics, University of Colorado,
Boulder, CO 80309-0440}
\author{Chris H. Greene}
\affiliation{JILA and Department of Physics, University of Colorado,
Boulder, CO 80309-0440}
\author{D. Blume}
\affiliation{JILA and Department of Physics, University of Colorado,
Boulder, CO 80309-0440}
\affiliation{Department of Physics and Astronomy,
Washington State University,
  Pullman, Washington 99164-2814}

\date{\today}

\begin{abstract}
We determine the energetically lowest lying states in the BEC-BCS
crossover regime of $s$-wave interacting two-component Fermi gases
under harmonic confinement by solving the many-body Schr\"odinger
equation using two distinct approaches.
Essentially exact basis set expansion
techniques are applied to determine the energy spectrum of
systems
with $N=4$ fermions. Fixed-node
diffusion Monte Carlo methods are applied to
systems with up to $N=20$ fermions, and
a discussion of different guiding functions used in the Monte Carlo approach
to impose the
proper symmetry of the fermionic system is presented.
The energies are calculated as a function of the $s$-wave scattering length
$a_s$ for $N=2-20$ fermions and different mass ratios $\kappa$ of
the two species. On the BEC and BCS sides, our energies agree with
analytically-determined first-order correction terms. We extract the
scattering length and the effective range of
the dimer-dimer system
up to $\kappa = 20$.
Our energies for the strongly-interacting trapped system in the unitarity
regime show no shell structure, and
are well described by a simple expression,
whose functional form can be derived using the local density approximation,
with one or two parameters.
The
universal parameter $\xi$ for the trapped system for various $\kappa$
is determined, and
comparisons with results for
the homogeneous system are presented.
\end{abstract}


\maketitle

\section{Introduction}
\label{introduction} Advances in trapping and cooling have spawned
the experimental realization of ultracold externally-confined
two-component Fermi gases with controllable interaction strengths.
Using these impurity-free systems, the crossover from a
weakly-attractive atomic Fermi gas through a strongly-interacting
unitarity regime to a weakly-repulsive molecular Bose gas has been
investigated~\cite{rega03a,zwie03,stre03,joch03}. Our increased
understanding of these systems relates to the study of neutron
matter and potentially that of high-T$_c$ superconductors, in
addition to the field of ultracold atomic gases. All of these
systems are controlled by similar pairing mechanisms, although at
much different densities.

To date, experimental studies of the BEC-BCS crossover with
ultracold atomic gases have been restricted to fermions in different
hyperfine substates. In this case, the ``spin-up'' and ``spin-down''
fermions have equal masses and experience equal trapping
frequencies. Currently, the simultaneous trapping and cooling of
different atomic fermionic species is being pursued in a number of
laboratories. This motivates us to investigate how the BEC-BCS
crossover physics changes with the mass ratio $\kappa$ of the two
atomic species. Our goal is to develop a microscopic understanding
of these intricate many-body systems. To this end, we consider
trapped systems with varying number of particles $N$, and relate
them to the homogeneous systems through the local density
approximation (LDA). This illuminates the transition from the
few-body to the many-body physics of an ultracold Fermi gas.

For a given short-range two-body potential, the stationary
Schr\"odinger equation for a trapped two-component Fermi gas has a
rich eigenspectrum, which in some cases includes deeply- and/or
weakly-bound cluster states as well as ``ground'' and highly-excited
gas-like states. In general, the eigenstates of two-component Fermi
gases with short-range interactions can be separated into two
classes: universal states that do not (or only weakly) depend on the
details of the two-body potential~\cite{footnote}, and non-universal states that
depend notably on the details of the two-body potential. The
eigenstates of the four-fermion system with equal masses, e.g., fall
into the former class, provided the range of the two-body potential
is sufficiently small; in this case, the properties of the system
are to a very good approximation determined by a single parameter,
the $s$-wave scattering length $a_s$. For large mass ratios,
however, non-universal bound trimer states
exist~\cite{efim70,efim73,petr05}. A description of these states
requires a three-body parameter, which depends on the short-range
physics. In some cases, non-universal bound clusters consisting of
four or more fermions may exist. In this work, we do not analyze the
properties of such non-universal states but instead study the
properties of states that depend at most weakly on the short-range
physics. In particular, we determine the BEC-BCS energy crossover
curve, which is defined in Sec.~\ref{crossover}, by solving the
stationary Schr\"odinger equation for various mass ratios $\kappa$.
An analysis of the stability of two-component Fermi systems with
large mass ratios, including molecular Bose gases created from
two-component Fermi gases, is beyond the scope of this paper.

Consider the stationary solutions of the four-particle system as a
function of $\kappa$ in the BEC-BCS crossover. The Schr\"odinger
equation is solved in two distinct approaches: a basis set expansion
technique that utilizes correlated Gaussians (CG) and a fixed-node
diffusion Monte Carlo (FN-DMC) approach. The dimer-dimer scattering
length $a_{dd}$ and the dimer-dimer effective range $r_{dd}$ emerge
as a function of $\kappa$. A surprisingly large $r_{dd}$ is found,
which is likely to be an important input parameter in the BEC
many-body theory. Furthermore, a detailed comparison of the results
throughout the entire crossover regime permits a non-trivial test of
the nodal surface employed in the FN-DMC approach, and it conveys
information about the symmetry of the many-body wave function.
Extension of our FN-DMC calculations to larger numbers of particles
also probes the validity range of the analytically-determined
limiting behaviors in the deep BCS and BEC regimes. In the
strongly-interacting unitarity regime, the LDA relates the trapped
system properties to those of the homogeneous system. Finally, our
FN-DMC energies should allow for a stringent test of numerically
less involved approaches such as density functional
treatments~\cite{bulgac07}.

Section~\ref{system} introduces the Hamiltonian of the trapped Fermi
system, and the numerical approaches applied to solve the
corresponding stationary many-body Schr\"odinger equation.
Section~\ref{results} presents our results for
%and interpretation of
 the energetics and the interpretation of the results of
weakly- and strongly-interacting Fermi systems with up to
$20$ atoms. Finally, Sec.~\ref{conclusion} concludes.

\section{Hamiltonian and numerical approach}
\label{system}

\subsection{Hamiltonian}
\label{hamiltonian}
For $N$ harmonically-trapped
Fermi atoms divided equally into
two species, the Hamiltonian
is given by
\begin{eqnarray}
\label{eq_ham}
H =
\sum_{i=1}^{N/2} \left(\frac{-\hbar^2}{2m_1} \nabla_i^2 +
\frac{1}{2} m_1 \omega_1^2 \vec{r}_i^2 \right) +
\nonumber \\
\sum_{i'=1}^{N/2} \left( \frac{-\hbar^2}{2m_2} \nabla_{i'}^2 +
\frac{1}{2} m_2 \omega_2^2 \vec{r}_{i'}^2 \right)
+\sum_{i=1}^{N/2} \sum_{i'=1}^{N/2} V(r_{ii'}).
\end{eqnarray}
Here, unprimed indices label mass $m_1$ and primed indices mass
$m_2$ fermions, and $N$ is assumed to be even. The mass ratio
$\kappa$ is defined by $m_1/m_2$ and throughout we take $m_1 \ge
m_2$. In Eq.~(\ref{eq_ham}), $\omega_1$ and $\omega_2$ denote
angular trapping frequencies, and $\vec{r}_i$ the position vector of
the $i$th fermion.

We adopt two purely attractive short-range model potentials for the
interaction between unlike fermions: a Gaussian interaction
potential $V(r)$, $V(r)=-V_0 \exp(-r^2/(2 R_0^2))$, and a square
well interaction potential $V(r)$, $V(r) = -V_0$ for $r < R_0$ and 0
otherwise. For a fixed range $R_0$ of the two-body potential $V(r)$,
the depth $V_0$, $V_0 \ge 0$, is adjusted until the $s$-wave
scattering length $a_s$ assumes the desired value. For negative (or
positive) $a_s$, $V_0$ and $R_0$ are chosen so the potential
supports no (or one) two-body $s$-wave bound state. Throughout this
paper, we treat the like atoms as non-interacting. This is justified
because the interactions between like atoms are only non-negligible
very close to a $p$-wave Feshback resonance. All experiments to date
have studied the BEC-BCS crossover using magnetic field strengths
for which the $p$-wave interactions are non-resonant.

The Hamiltonian given by Eq.~(\ref{eq_ham}) is characterized by four
different length scales: the range $R_0$ of the interaction
potential, the $s$-wave scattering length $a_s$, and the two
oscillator lengths $a_{ho}^{(j)} = \sqrt{\hbar/(m_j \omega_j)}$,
$j=1$ and $2$. Throughout, we are interested in the regime where
$R_0$ is much smaller than the oscillator lengths $a_{ho}^{(j)}$, or
equivalently, where the system is dilute with respect to $R_0$,
i.e., $n(0) R_0^3 \ll 1$, where $n(0)$ denotes the peak density. In
this regime, the properties of the universal states are expected to
be independent of the details of the two-body potential. For a given
$a_s$, we numerically test whether a state is universal by
calculating its energy for various $R_0$. The condition $R_0 \ll
a_{ho}^{(j)}$ implies that the numerical approaches chosen for
solving the many-body Schr\"odinger equation have to be able to
govern the physics occuring at at least two different length scales.
As we illustrate below, the CG and FN-DMC approaches are able to do
so.

\subsection{Correlated Gaussian approach}
\label{cgmethod} The correlated Gaussian
method\cite{singer1960uge,suzuki1998sva} is a powerful tool to study
few-body systems. Recently, the CG approach has been applied to the
four-fermion system with equal masses~\cite{stec07}. Here, we
analyze the properties of this system from a somewhat different
point of view and additionally consider the unequal mass system. As
in the previous work, we treat equal trapping frequencies, i.e.,
$\omega_1=\omega_2$, so that the center-of-mass motion separates. To
further reduce the dimensionality of the problem, we restrict
ourselves
 to states with vanishing total angular momentum $L$
and positive parity $P$.
We expand the $L^P=0^+$ states in terms of correlated Gaussian basis
functions $\Phi_{\vec{d}}$, which depend on the six
interparticle distances and the center of mass vector,
\begin{equation}
\psi(\vec{r}_1,\vec{r}_2,\vec{r}_{1'},\vec{r}_{2'})=
\sum_{\{\vec{d}\}} C_{\vec{d}}\,\Phi _{\vec{d}}.
  \label{TotalWF}
\end{equation}%
Here, the $C_{\vec{d}}$ denote expansion coefficients. Each basis
function $\Phi _{\vec{d}}$ is written as a product of the ground
state center-of-mass wavefunction and a symmetrized product of
Gaussian functions for each interparticle distance
vector~\cite{stec07}. The widths of these Gaussians can be different
for each interparticle distance, giving us six parameters for each
basis function. These parameters are in Eq.~({\ref{TotalWF})
collectively denoted by $\vec{d}$. The simple functional form of the
wave function $\psi$, Eq.~(\ref{TotalWF}), allows the analytical
determination of all matrix elements if the two-body interaction
potential is taken to be a Gaussian (see Sec.~\ref{hamiltonian}).

The parameter vector $\vec{d}$ that characterizes each basis
function $\Phi_{\vec{d}}$ are selected semi-randomly. Typically, the
components of $\vec{d}$ vary from a fraction of the range $R_0$ to a
few times the interparticle distance in the noninteracting limit.
The basis functions can be roughly separated into three types. The
first type has all the components of $\vec{d}$ of the order of the
trap lengths and is suitable to describe gas-like states. The second
type has one or two small $\vec{d}$ components while the others take
values of the order of the trap lengths; these basis functions carry
a large weight when $\psi$ describes states that consist---to a good
approximation---of two bound dimers or a dimer and two free atoms.
The third type has more than two small $\vec{d}$ components, and is
suitable to describe comparatively tightly-bound three- and
four-body states. In general, all three types of basis functions are
needed to accurately describe the entire $L^P=0^+$ spectrum of the
four-fermion system. For equal masses, however, we find that the
third type carries negligible weight, owing to the absence of
molecular three- or four-body states. As the mass ratio increases,
more basis functions of the third type need to be included. We
carefully check the convergence of the energies by varying the total
number of basis functions used, and by varying the basis functions
included in the expansion. We find that of the order of $10^4$ basis
functions suffice to accurately describe the eigenfunctions of
interest.

The basis functions introduced above are not linearly independent.
To eliminate the linear dependence in our basis set, we diagonalize the
overlap matrix and eliminate the basis functions with the lowest
eigenvalues up to a certain cutoff. The remaining basis functions
are then used to construct
a new orthogonal basis set.
Finally, the eigenspectrum is obtained by diagonalizing
the corresponding Hamiltonian matrix.

\subsection{Fixed-node diffusion Monte Carlo approach}
\label{fndmcmethod}

To treat up to $N=20$ fermions, solutions of the Schr\"odinger
equation are determined by the FN-DMC method~\cite{reyn82,hamm94}.
In this method, the proper fermionic antisymmetry is imposed through
the use of a so-called guiding function $\psi_T$, which depends on
the coordinates of all particles. To within statistical
uncertainties, the FN-DMC algorithm provides an upper bound to the
exact ground state energy, i.e., to the lowest-lying state with the
same symmetry as $\psi_T$. Note that all our FN-DMC calculations are
performed for the square well interaction potential. While there is
no technical problem in extending the FN-DMC calculations to the
Gaussian interaction potential, the guiding functions $\psi_T$ are
most readily determined and evaluated for the square well potential.

Two different guiding functions $\psi_T$ considered in this work are:
\begin{eqnarray}
\label{eq_t1}
\psi_{T1}=
%\left[
%\prod_{i=1}^{N/2} \exp\left(-\frac{1}{2}(\frac{\vec{r}_i}{b_M})^2\right)
%\right]
%\left[
%\prod_{i'=1}^{N/2} \exp\left(-\frac{1}{2} (\frac{\vec{r}_{i'}}{a_{ho}^{(2)}})^2 \right)
%\right]
\prod_{i=1}^{N/2} \Phi(\vec{r}_i/a_{ho}^{(1)})
\times
\prod_{i'=1}^{N/2} \Phi(\vec{r}_{i'}/a_{ho}^{(2)}) \times
\nonumber \\
{\cal{A}}
(f(r_{11'}),\cdots, f(r_{N/2,N/2'}))
\end{eqnarray}
and
\begin{eqnarray}
\label{eq_t2}
\psi_{T2}=
\Psi_{NI}(\vec{r}_1,\cdots,\vec{r}_{N/2'}) \times
\prod_{i,i'}^{N/2} \bar{f}(r_{ii'}),
%Det( \Phi_1(\vec{r}_{1}),\cdots, \Phi_{N/2}(\vec{r}_{N/2})) \nonumber \\
%Det( \Phi_1(\vec{r}_{1'}),\cdots, \Phi_{N/2}(\vec{r}_{N/2'})).
\end{eqnarray}
Here $\Phi$ denotes the ground state harmonic oscillator orbital,
${\cal{A}}$ is the antisymmetrizer, and $\Psi_{NI}$ denotes the wave
function of $N$ trapped non-interacting fermions. Following
Ref.~\cite{astr04c}, the pair function $f$ is constructed from the
free-space zero-energy scattering and the free-space bound state
solution of the two-body square-well interaction potential for
negative and positive $s$-wave scattering length $a_s$,
respectively. In Eq.~(\ref{eq_t2}), $\bar{f}$ coincides with $f$ for
small $r$ and is matched smoothly to a non-zero constant at larger
$r$. This matching to a non-zero constant ensures that the product
over all pair functions $\bar{f}$ is always non-zero. Thus, the
nodal structure of $\psi_{T2}$ coincides with that of the
non-interacting Fermi gas. In contrast, the nodal surface of
$\psi_{T1}$ is constructed by antisymmetrizing a product of pair
functions~\cite{astr04c}.

To assess the accuracy of our MC code, we determine the energy of
the two-body system with $\omega_1=\omega_2$ and $m_1=m_2$
semi-analytically. We separate off the center-of-mass motion, and
write the eigenfunctions of the Schr\"odinger equation for the
relative coordinate in terms of hypergeometric functions. The
resulting eigenequation results in an energy of $E=2.00200 \hbar
\bar{\omega}$.
Since the two-body wave function is
nodeless, the DMC energy for $N=2$ (see Table~\ref{energies})
\begin{table}
\caption{\label{energies} FN-DMC energies $E$ in units of $\hbar
\bar{\omega}$ as a function of $N$, $N=2-20$, for the two-component
Fermi system with $a_{ho}^{(1)} = a_{ho}^{(2)}$ at unitarity for
$\kappa=1$ and 8. Statistical uncertainties of the energies are
reported in round brackets. Note that no nodal approximation needs
to be made for $N=2$.}
\begin{ruledtabular}
\begin{tabular}{cll}
 $N$ & $E(N)/(\hbar \bar{\omega})$ ($\kappa=1$) & $E(N)/(\hbar \bar{\omega})$ ($\kappa=8$) \\ \hline
2 &  2.00202(3) & 1.726(7) \\
4 &  5.069(9)  & 4.48(1) \\
6 &  8.67(3)  & 7.82(3) \\
8 &  12.57(3)  & 11.39(4) \\
10 & 16.79(4)  & 15.28(6) \\
12 & 21.26(5)  & 19.45(6) \\
14 & 25.90(5) & 23.81(6) \\
16 & 30.92(6)  & 28.46(7) \\
18 & 36.00(7)  & 33.42(8) \\
20 & 41.35(8)  & 38.11(9) \\
\end{tabular}
\end{ruledtabular}
\end{table}
is expected to be exact. Indeed, the DMC
energy agrees to within the statistical
uncertainty with the energy determined semi-analytically.
A detailed comparison of the FN-DMC and CG energies for the four-fermion
system, which allows the quality of the nodal surface employed
in the FN-DMC calculations to be assessed, is presented in
Sec.~\ref{fourbody}.

In the non-interacting case, i.e., for $a_s=0$, the guiding function
$\psi_{T2}$ with $\bar{f}(r)=1$ coincides with the exact eigen
function. For weakly-attractive Fermi systems, the attractive nature
of the two-body potential introduces correlations but does, to a
good approximation, leave the nodal surface unchanged. Indeed, we
find that the variational energy for $\psi_{T1}$ in this regime is
nearly indistinguishable from the FN-DMC energies, indicating that
the Jastrow product over all pair functions accounts properly for
the two-body correlations of the system and that three- and
higher-order correlations are negligible.

For small positive $a_s$, on the other hand, comparatively
strongly-bound two-body dimers exist and the system is expected to
form a molecular Bose gas of dimers. Such a system is not even
qualitatively described correctly by the guiding function
$\psi_{T1}$, which assumes that every spin-up fermion is
``simultaneously'' correlated with every spin-down
fermion~\cite{footnote4}. The guiding function $\psi_{T2}$, instead,
is much better suited to describe a Fermi gas that behaves as a
weakly-interacting molecular Bose gas. $\psi_{T2}$ correlates the
first spin-up fermion with the first spin-down fermion, the second
spin-up fermion with the second spin-down fermion, and so on, and
then anti-symmetrizes this ``paired state''. The guiding function
$\psi_{T2}$ is expected to accurately describe the system when the
size of the dimer pairs becomes small compared to the oscillator
lengths.

Finally, in the strongly-interacting regime, i.e., for $|a_s|
\rightarrow \infty$, it is not {\em{a priori}} clear which of the
two guiding functions provides a better description of the system.
Section~\ref{results} discusses this in more detail, and also
comments on additional aspects of the choice of the guiding
functions related to the existence of non-universal trimer states.

\section{Results}
\label{results}

\subsection{Energy crossover curve: Definition and general considerations}
\label{crossover} Throughout this work, we are interested in
describing the crossover from a weakly-repulsive to a
weakly-attractive trapped two-component Fermi system with $N$
particles. This BEC-BCS crossover can be characterized by the
normalized energy crossover curve $\Lambda_N^{(\kappa)}$,
\begin{eqnarray}
\label{eq_cross}
\Lambda_N^{(\kappa)} =
\frac{E(N)-NE(2)/2}{\lambda \hbar \bar{\omega}},
\end{eqnarray}
which depends on $a_s$, $\kappa$ and $N$. In Eq.~(\ref{eq_cross}),
$E(N)$ denotes the energy of the $N$-fermion system, $\bar{\omega} =
(\omega_1+\omega_2)/2$ is the average frequency and $\lambda$ is
defined through the energy $E_{NI}$ of $N$ non-interacting fermions,
\begin{eqnarray}
\label{eq_eni}
E_{NI}=(\lambda + 3N/2) \hbar \bar{\omega}.
\end{eqnarray}
The values of $\lambda$ for the first few closed-shell systems are
listed in the second column of Table~\ref{Ckappas}.
$\Lambda_N^{(\kappa)}$ equals one on the deep BCS side (small
$|a_s|$ and $a_s<0$), and zero on the deep BEC side (small $a_s$ and
$a_s>0$). Since the energy $E(2)$ of $N/2$ trapped dimer pairs is
subtracted from the total energy $E(N)$ of the system, the energy
crossover curve $\Lambda_N^{(\kappa)}$, Eq.~(\ref{eq_cross}), is
expected to be independent of the details of the two-body potential
if the range $R_0$ is much smaller than the average interparticle
spacing. The energy crossover curve defined here for the trapped
system is the analog of the BEC-BCS crossover curve of the
homogeneous system (see, e.g., Refs.~\cite{legg80,nozi85,enge97} for
pioneering work based on the mean-field BCS equations, and Figs.~1
and 2 of Ref.~\cite{astr04c} for a determination of the crossover
curve for the homogeneous system by the FN-DMC method).

As indicated above, $\Lambda_N^{(\kappa)}$
depends on the scattering length $a_s$, the number of
particles $N$, the ratio between the two masses, and the
ratio between the two frequencies. Thus, an exhaustive
study of the whole parameter space of
trapped two-component Fermi systems by
first principle methods is
impossible.
This paper considers two different scenarios:
i) the trapping frequencies are set to coincide, i.e., $\omega_1=\omega_2$,
while $\kappa$, $N$ and $a_s$ are varied (see Secs.~\ref{fourbody} and
\ref{nbody}), and
ii) the oscillator lengths are set to coincide, i.e.,
$a_{ho}^{(1)}= a_{ho}^{(2)}$, and $1/|a_s|$ is set to 0, while
$N$ and $\kappa$ are varied (see Sec.~\ref{unitarity}).

Our motivation for considering scenario i) is as follows. In the
deep BEC regime, the fermionic system is expected to form a
molecular Bose gas whose behaviors are to a good approximation
determined by the dimer-dimer scattering length $a_{dd}$. The
dimer-dimer scattering length can be extracted quite readily for
different $\kappa$ from our four-body energies, provided the
center-of-mass motion decouples (see Sec.~\ref{fourbody}). For
unequal frequencies, the center-of-mass motion does not decouple and
the extraction of the dimer-dimer scattering length would be more
involved. Section~\ref{nbody} extends the study of the four-fermion
system with equal frequencies to systems with more particles to
illustrate that the behaviors of the larger systems in the deep BEC
regime are also governed to a good approximation by the dimer-dimer
scattering length. While this has been shown to be the case
previously for the homogeneous system with equal masses~\cite{astr04c}, our
calculations illustrate that---as might be expected---the many-body
physics of unequal mass systems in the deep BEC regime is also to a
good approximation governed by a single few-body parameter, the
dimer-dimer scattering length. In the more strongly-interacting
regime, our equal frequency calculations for $\kappa >1$ provide
insights into the behaviors of systems whose densities are not fully
overlapping. Thus, mass-imbalanced Fermi system may behave in certain
respects similar to population-imbalanced Fermi systems.

Our primary motivation for considering scenario ii) is to connect
the behaviors of the trapped system with those of the homogeneous
system using the LDA. The energy of the homogeneous system at
unitarity is related to the energy of the non-interacting system by
a universal parameter $\xi$. By calculating the energies of the
trapped system at unitarity for equal frequencies and equal masses,
we quantify how well the LDA describes small trapped systems.
Unpublished results for the homogeneous system with equal spin-up
and spin-down densities but unequal masses indicate that the
universal parameter $\xi$ depends weakly on the mass ratio
$\kappa$~\cite{astr07}. These unequal mass results for the
homogeneous system cannot be straightforwardly connected to those of
the trapped system since the densities of the two trapped species do
not necessarily fully overlap. In the non-interacting limit, the
densities of the two species coincide when the trapping lengths are
equal, i.e., $a_{ho}^{(1)} = a_{ho}^{(2)}$. At unitarity, we find
that unequal mass systems with $a_{ho}^{(1)} = a_{ho}^{(2)}$ have a
small, though non-negligible, density mismatch. Despite this small
density mismatch, we apply the LDA $a_{ho}^{(1)} = a_{ho}^{(2)}$ and
relate the universal parameter of the homogeneous system to the
energetics of the trapped system.

\subsection{Energy crossover curve for $N=4$}
\label{fourbody}
This  subsection presents our results for the
energy crossover curve for four trapped fermions calculated by the
CG and FN-DMC approaches for equal frequencies, i.e.,
$\omega_1=\omega_2=\bar{\omega}$, and varying mass ratio $\kappa$.

Figure~\ref{fig1} shows the energy crossover curve
\begin{figure}
\vspace*{.075in} \centerline{\epsfxsize=3.0in\epsfbox{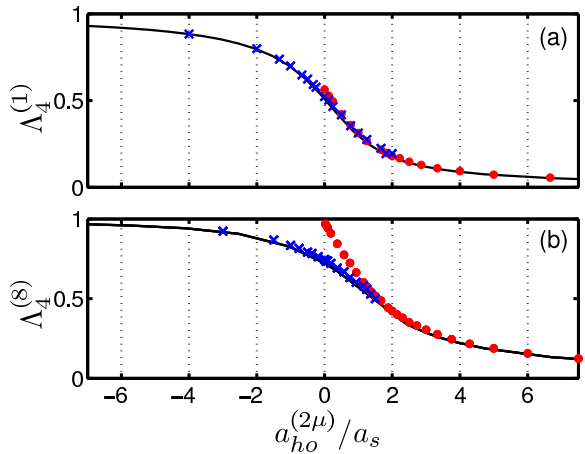}}
%\vspace*{-1.6in}
\caption{ (Color online) Energy crossover curve
$\Lambda_4^{(\kappa)}$ for $\omega_1=\omega_2$ as a function of
$a_{ho}^{(2\mu)}/a_s$ for (a) $\kappa=1$ and (b)  $\kappa=8$. Solid
lines are calculated by the CG approach, and circles and crosses by
the FN-DMC method using $\psi_{T1}$ and $\psi_{T2}$, respectively.}
\label{fig1}
\end{figure}
$\Lambda_4^{(\kappa)}$  for four fermions as a function of
$a_{ho}^{(2 \mu)}/a_s$ calculated by the CG and FN-DMC approaches
for (a) $\kappa=1$ and (b) $\kappa=8$. Here, the oscillator length
$a_{ho}^{(2 \mu)}$ is defined in terms of the reduced mass
$\mu=m_1m_2/(m_1+m_2)$, i.e., $a_{ho}^{(2 \mu)} = \sqrt{\hbar/(2 \mu
\bar{\omega})}$. The solid lines in Fig.~\ref{fig1} are obtained
using $E(4)$ calculated by the CG approach, while circles and
crosses are obtained using $E(4)$ calculated by the FN-DMC approach
using $\psi_{T1}$ and $\psi_{T2}$, respectively. The ranges $R_0$ of
the two-body potentials used in Fig.~\ref{fig1} are much smaller
than the oscillator lengths, i.e., $R_0 \approx 0.01a_{ho}^{(2
\mu)}$. From our CG energies for different $R_0$, we estimate that
the $\Lambda_N^{(\kappa)}$ shown in Fig.~\ref{fig1} deviate by at
most 1\% from the corresponding curves for zero-range interactions.
For $m_1=m_2$, e.g., the energy at unitarity calculated by the CG
approach for the Gaussian interaction potential is $E = 5.027 \hbar
\bar{\omega}$ for $R_0=0.01 a_{ho}^{(2\mu)}$ and $E = 5.099 \hbar
\bar{\omega}$ for $R_0=0.05 a_{ho}^{(2\mu)}$~[5]. For comparison,
the FN-DMC energy for the square well potential with $R_0=0.01
a_{ho}^{(2 \mu)}$ is $E=5.069(9)$ (see Table~\ref{energies}), which
is in good agreement with the energies calculated by the CG
approach.

As expected, the energy crossover curve connects the limiting values
of one on the BCS side and zero on the BEC side smoothly.
Importantly, the lowest FN-DMC energies and the CG energies agree
well, which implies that the functional forms of $\psi_{T1}$ and
$\psi_{T2}$ are adequate. For equal masses [panel (a)], the FN-DMC
energies at unitarity calculated using the two different $\psi_T$
agree approximately. For unequal masses [panel (b)], in contrast,
the nodal surface of $\psi_{T2}$ leads to a lower energy at
unitarity than that of $\psi_{T1}$, and the crossing point between
the energies calculated using $\psi_{T1}$ and $\psi_{T2}$ moves to
the BEC side. This can be understood by realizing that the densities
of the heavy and light particles do not overlap fully, leading to a
reduced pairing.

The CG approach in our current implementation (see Sec.~\ref{cgmethod})
allows for the determination of the complete
$L^P=0^+$ energy spectrum. If we use short-range Gaussian two-body
potentials that support no two-body $s$-wave bound state for
negative $a_s$ and one two-body $s$-wave bound state for positive
$a_s$, the four-body energy that enters the calculation of the
energy crossover curves shown in Fig.~\ref{fig1} is the true ground
state of the system, i.e., no energetically lower-lying bound trimer or
tetramer states with $L^P=0^+$ symmetry exist. For larger mass
ratios, bound trimer states exist. The mass ratio
at which these non-universal trimer states
appear depends on the range $R_0$ of the two-body potential
employed. In the regime where three-body bound states exist, the four-body
spectrum calculated by the CG approach contains also universal
states which are separated by approximately $2 \hbar \bar{\omega}$
and which can be best described as two weakly-interacting composite
bosons. For fixed $a_s$, $a_s > 0$, the energy of these
``dimer-dimer states'' changes smoothly as a function of $\kappa$
even in the regime where bound trimer states appear. In the
following, we use these dimer-dimer states to extract the
dimer-dimer scattering length as a function of $\kappa$ up to
$\kappa = 20$.

When $a_s$ is small, the four fermions form two bosonic molecules of
mass $M$, where $M=m_1+m_2$, which interact through an effective
molecule-molecule potential~\cite{petr05}. To model this effective
dimer-dimer potential (the exact functional form is unknown), we
introduce a regularized zero-range potential $V(r)$~\cite{huan57},
$V(r)=g \delta(\vec{r})
(\partial/\partial r) r$, whose scattering strength $g$ is
parameterized by the scattering length $a_{dd}$ and the effective
range $r_{dd}$ of the dimer-dimer system, i.e.,
\begin{eqnarray}
g=\frac{4 \pi \hbar^2 \, a_{dd}}
{M}
\left[1- \frac{M E_{tb} r_{dd}a_{dd}}{2 \hbar^2}
\right]^{-1}.
\end{eqnarray}
Here, $E_{tb}$ denotes the relative energy of the two-boson system,
i.e., the total energy with the center-of-mass contribution
subtracted. It has been shown previously~\cite{blum02,bold02} that
the inclusion of the energy dependence of the scattering length
notably extends the validity regime of the zero-range
pseudopotential when applied to describe the scattering of two atoms
under external confinement. By comparing the ``dimer-dimer energy
levels'' of the four-fermion system with the energies of two mass
$M$ bosons in a trap interacting through this energy-dependent
zero-range potential~\cite{busc98,blum02,bold02}, we determine
$a_{dd}$ and $r_{dd}$.

\begin{figure}
\vspace*{.075in} \centerline{\epsfxsize=3in\epsfbox{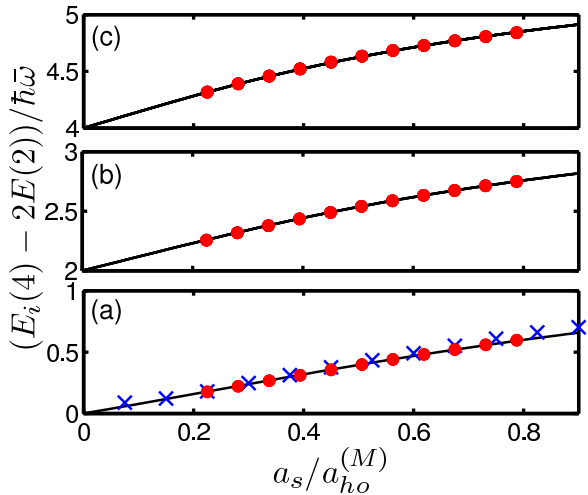}}
%\vspace*{-1.6in}
\caption{ (Color online) Four-body energies of the three
energetically lowest-lying dimer-dimer states as a function of
$a_s/a_{ho}^{(M)}$ for $\kappa=8$. Panel~(a) shows the energetically
lowest lying energy level ($i=0$), panel (b) the energetically
second lowest ($i=1$) and panel (c) the energetically third lowest
state ($i=2$). Circles and crosses show our CG and FN-DMC results,
respectively. Solid lines show the zero-range model results.}
\label{spectrum}
\end{figure}

To illustrate this procedure,
circles in Figs.~\ref{spectrum}(a) through (c) show the three
energetically lowest-lying dimer-dimer energy levels, referred to
as $E_i(4)$ with $i=0$ through $2$,
with the center-of-mass energy and the dimer-binding energy
subtracted for $\kappa=8$
 obtained by the
CG approach as a function of $a_s$.
Solid
lines show the energy levels obtained by
fitting these four-body energies by the two-boson energies
obtained using the energy-dependent zero-range
pseudopotential ($a_{dd}$ and $r_{dd}$ are treated as fitting parameters).
We find that inclusion of the effective range $r_{dd}$
extends the validity regime over which the
four-fermion system can be described by the two-boson model
and additionally allows for a more reliable determination
of $a_{dd}$.
Figure~\ref{spectrum} illustrates that the two-boson spectrum
reproduces the dimer-dimer states
of the four-fermion spectrum well over a fairly large
range of atom-atom scattering lengths $a_s$.
For comparison, crosses in Fig.~\ref{spectrum}(a)
show the corresponding FN-DMC energies for the energetically
lowest-lying dimer-dimer state. We did not attempt to construct a
guiding function that would allow for the determination of excited
dimer-dimer states. We find that the FN-DMC energies are slightly
larger than the CG energies and that the deviation increases with
increasing $a_s$. Presumably, this can be attributed to the
functional form of the nodal surface used in the FN-DMC
calculations, which should be best in the very deep BEC regime. The
increasing deviation between the FN-DMC and CG energies with
increasing $a_s$ explains why the effective range predicted by the
analysis of the FN-DMC energies is somewhat larger than that
predicted by the analysis of the CG approach (see discussion of
Fig.~\ref{fig2} below).

Circles and crosses
in Fig.~\ref{fig2} show the resulting dimer-dimer scattering
\begin{figure}
\vspace*{.075in} \centerline{\epsfxsize=3.in\epsfbox{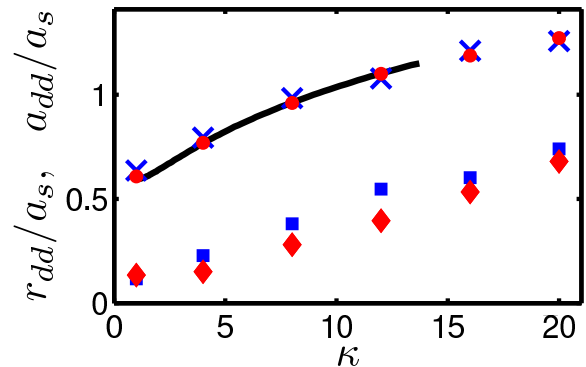}}
%\vspace*{-1.6in}
\caption{ (Color online) Circles and crosses show $a_{dd}/a_s$ as a
function of $\kappa$ extracted from the four-fermion CG and FN-DMC
energies, respectively. For comparison, a solid line shows the
results from Fig.~3 of Ref.~\protect\cite{petr05}. Diamonds and
squares show $r_{dd}/a_s$ extracted from the four-fermion CG and
FN-DMC energies, respectively.} \label{fig2}
\end{figure}
lengths $a_{dd}$ extracted from the energies calculated by the CG
and the FN-DMC approach, respectively, as a function of $\kappa$.
For all mass ratios considered in Fig.~\ref{fig2}, we include up to
three dimer-dimer energy levels in our analysis of the CG results,
and only the lowest dimer-dimer level in our analysis of the FN-DMC
results. Our dimer-dimer scattering lengths agree well with those
calculated by Petrov {\em{et al.}} within a zero-range
framework~\cite{petr05} (solid line in Fig.~\ref{fig2}). The
calculations by Petrov~{\em{et al.}},
performed for the free 
and not 
the trapped four-fermion system, terminate at $\kappa \approx
13.6$, beyond which a three-body parameter is needed to solve the
four-body equations within the applied framework. 
Our calculations show the existence of deeply-bound ``plunging'' states,
which 
consist of
a trimer plus a free atom.
This signals a qualitative change of the energy 
spectrum, in agreement with Petrov {\em{et al.}}~\cite{petr05}.
At the same time, our calculations
for finite-range potentials predict that $a_{dd}$ continues to
increase smoothly when the mass ratio $\kappa$ exceeds 13.6. 
This can possibly be explained by the fact that
the presence of the external confining potential may ``wash out'' 
some of the features
present in the free-space system. 
As
already mentioned, the study of the stability of the four-fermion
system, consisting of two dimers, 
with large mass ratios is beyond the scope of this work.

Diamonds and squares in Fig.~\ref{fig2} show the effective range
$r_{dd}$ extracted from our CG and FN-DMC energies, respectively. We
estimate the uncertainty of $r_{dd}$ obtained from the CG approach
to be about 10\%, and quite a bit larger for that extracted from the
FN-DMC energies. Figure~\ref{fig2} shows that the ratio
$r_{dd}/a_{dd}$ increases from about 0.2 for $\kappa=1$ to about 0.5
for $\kappa=20$. While earlier work already suggested that the
dimer-dimer potential may be best characterized as a broad soft-core
potential~\cite{petr05}, implying a non-negligible value for the
effective range $r_{dd}$, our work makes the first quantitative
predictions for $r_{dd}$ as a function of $\kappa$. The large value
of $r_{dd}$ suggests that effective range corrections may need to be
considered in analyzing the physics of molecular Fermi gases.

\subsection{Weakly-interacting limits for $N>4$}
\label{nbody} We now apply the FN-DMC method to larger systems with
$\omega_1=\omega_2=\bar{\omega}$, focussing on the deep BCS and BEC
regimes where $|a_s|$ is small. Figure~\ref{fig3}(a) shows the total
energy $E(N)$ for $N=8$
\begin{figure}
\vspace*{.075in} \centerline{\epsfxsize=3.4in\epsfbox{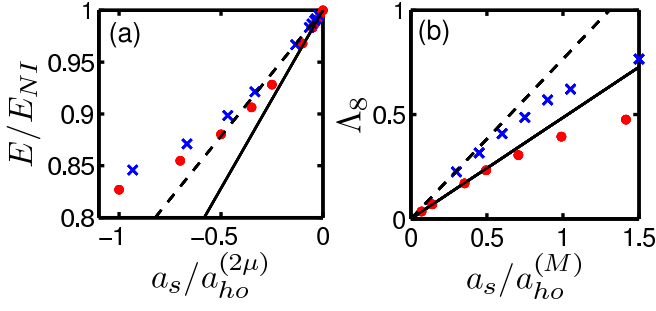}}
%\vspace*{-1.6in}
\caption{
(Color online)
Energies for $N=8$ and small $|a_s|$.
(a)
$E(8)/E_{NI}$ as a function of $a_s/a_{ho}^{(2 \mu)}$
for $\omega_1=\omega_2=\bar{\omega}$
for $\kappa=1$ (circles and solid line) and
$\kappa=8$ (crosses and dashed line).
Symbols are calculated by the FN-DMC method using
$\psi_{T2}$, and lines using the first order correction,
Eq.~(\protect\ref{eq_expbcs}).
(b)
$\Lambda_8^{(\kappa)}$ as a function of $a_s/a_{ho}^{(M)}$
for $\omega_1=\omega_2=\bar{\omega}$
for $\kappa=1$ (circles and solid line) and
$\kappa=8$ (crosses and dashed line).
Symbols are calculated by the FN-DMC method using
$\psi_{T1}$, and lines using
%using
the first order correction,
Eq.~(\protect\ref{eq_expbec}).
}
\label{fig3}
\end{figure}
particles, divided by the energy $E_{NI}$ of the non-interacting system,
for $\kappa=1$ and $8$
as a function
of $a_s/a_{ho}^{(2 \mu)}$ for small $|a_s|$. The FN-DMC energies,
calculated using $\psi_{T2}$, are shown by symbols (circles for $\kappa=1$
and crosses for $\kappa=8$).

For comparison, we calculate the energy of a weakly-attractive
closed-shell Fermi system with equal frequencies ($\bar{\omega}=
\omega_1 = \omega_2$) in first order perturbation theory. We assume
that the unlike fermions are interacting through the Fermi
pseudopotential $V_s(\vec{r},\vec{r}')= \frac{2 \pi \hbar^2
a_s}{\mu} \delta(\vec{r}-\vec{r}')$~\cite{fermi36}. Applying
perturbation theory to the non-interacting two-component Fermi gas
with unequal masses but equal frequencies, the energy becomes $E
\approx E_{NI} + E_{int}^{(1)}$, where
 the first order energy correction $E_{int}^{(1)}$
for closed-shell systems can be written as
\begin{equation}
  \label{Ener}
E^{(1)}_{int}=\frac{2 \pi a_s\hbar^2}{\mu} \int \rho_{m_1}(\vec{r})
\rho_{m_2}(\vec{r}) d \vec{r}.
\end{equation}
Here, $\rho_{m_i}$ denotes the density of a single non-interacting
\begin{table}
\caption{\label{Ckappas} Values of $\lambda$ and $C_N^{\kappa}$ for
the four smallest closed-shell two-component Fermi systems with
equal frequencies.}
\begin{ruledtabular}
\begin{tabular}{ccc}
 $N$ & $\lambda$ & $C_{N}^{\kappa}$\\
\hline 2 & 0 & $\sqrt{\frac{2}{\pi}}$\\
 8& 6 & $\sqrt{\frac{2}{\pi}}\frac{(4+23\kappa+4\kappa^2)}{(1+\kappa)^{2}}$\\
 20& 30 & $\sqrt{\frac{2}{\pi}} \frac{5
(20+50\kappa+249\kappa^2+50\kappa^3+20\kappa^4)}{4(1+\kappa)^{4}}$\\
40& 90 &$\sqrt{\frac{2}{\pi}} \frac{5
(40+450\kappa+306\kappa^2+2795\kappa^3+306\kappa^4+450\kappa^5+40\kappa^6)}{4(1+\kappa)^{6}}$\\
\end{tabular}
\end{ruledtabular}
\end{table}
mass $m_i$ component ($i=1$ and $2$).
 The integration in Eq.~(\ref{Ener}) can be performed analytically,
resulting in a simple expression for the first-order energy
correction,
\begin{equation}
  \label{Ener2}
E^{(1)}_{int}=\hbar\bar{\omega} C_N^{\kappa} \frac{a_s}
{a_{ho}^{(2\mu)}},
\end{equation}
where $C_N^{\kappa}$ denotes a constant that depends on $N$ and
$\kappa$. Altogether, we obtain
\begin{eqnarray}
\label{eq_expbcs} E \approx E_{NI}  + \hbar \bar{\omega} \,
C_N^{\kappa} \, \frac{a_s}{a_{ho}^{(2 \mu)}}.
\end{eqnarray}
The values of $C_N^{\kappa}$ for the first four closed-shell systems
are summarized in the third column of Table~\ref{Ckappas}. For
completeness, the second column summarizes the values of $\lambda$
that determine the
energy $E_{NI}$ of the non-interacting system [see Eq.~(\ref{eq_eni})].

The first order correction,
Eq.~(\ref{eq_expbcs}), is shown by solid and dashed lines in
Fig.~\ref{fig3}(a) for $\kappa=1$ and 8, respectively; it describes
the interacting system well for $|a_s| \le 0.25 a_{ho}^{(2 \mu)}$,
or equivalently, for $k_F^0 a_s \le 0.6$. Here, $k_F^0$ denotes the
wave vector at the trap center of a non-interacting system of $N$
fermions with mass $2 \mu$ evaluated within the Thomas Fermi
approximation, $k_F^0=\sqrt{2} (3N)^{1/6}/a_{ho}^{(2 \mu)}$.
Additional corrections can be derived within a renormalized
scattering length framework~\cite{vonstech07}.

We now turn to the small $a_s$ limit for $N=8$.
Circles ($\kappa=1$) and crosses ($\kappa=8$) in Fig.~\ref{fig3}(b)
show the energy crossover curve $\Lambda_8^{(\kappa)}$ in the BEC regime for
$\bar{\omega} = \omega_1 = \omega_2$, where $E(8)$ is calculated by
the FN-DMC method, as a function of $a_s/a_{ho}^{(M)}$, where
$a_{ho}^{M}=\sqrt{\hbar/(M \bar{\omega})}$;
in
the small $a_s$ regime, $a_{ho}^{(M)}$ is the relevant characteristic
oscillator length. Treating the $N$-fermion system as a bosonic gas
consisting of $N/2$ mass $M$ molecules and applying first order
perturbation theory using a Fermi
pseudopotential~\cite{fermi36}, the energy of the system with
$\bar{\omega} = \omega_1 = \omega_2$ reads
\begin{eqnarray}
\label{eq_expbec} E \approx \frac{N}{2}E(2) + \hbar \bar{\omega} \,
\frac{N(N-2)}{8} \sqrt{\frac{2}{\pi}} \,
\frac{a_{dd}}{a_{ho}^{(M)}}.
\end{eqnarray}
Solid and dashed lines in Fig.~\ref{fig3}(b) show
$\Lambda_8^{(\kappa)}$ for $\kappa=1$ and $8$, respectively,
calculated using Eq.~(\ref{eq_expbec}). To plot the expansion, we
use the dimer-dimer scattering length $a_{dd}$ calculated by the CG
approach. For both mass ratios, the agreement between the FN-DMC
energies and the first order correction is good for $a_s \le 0.5
a_{ho}^{(M)}$. Figure~\ref{fig3}(a) thus illustrates that the
behavior of the Fermi system depends to a first approximation only
on $a_{dd}$ if $a_s/a_{ho}^{(M)}$ is sufficiently small. Inclusion
of effective range corrections may improve the agreement but is
beyond the scope of this paper.

We checked that the behaviors discussed here for $N=8$ also hold for
$N=20$ particles.

\subsection{Energetics at unitarity}
\label{unitarity} This section considers the strongly-interacting
unitary regime, where the atom-atom scattering length is infinite.
To ensure large overlap of the densities of the two species, we
choose the trapping frequencies $\omega_1$ and $\omega_2$  so that
$a_{ho}^{(1)}=a_{ho}^{(2)}$ (see Sec.~\ref{crossover} for a
discussion). Circles and crosses in Fig.~\ref{fig4} show our FN-DMC
energies $E(N)$ at unitarity as a function of $N$
\begin{figure}
\vspace*{.075in} \centerline{\epsfxsize=3.in\epsfbox{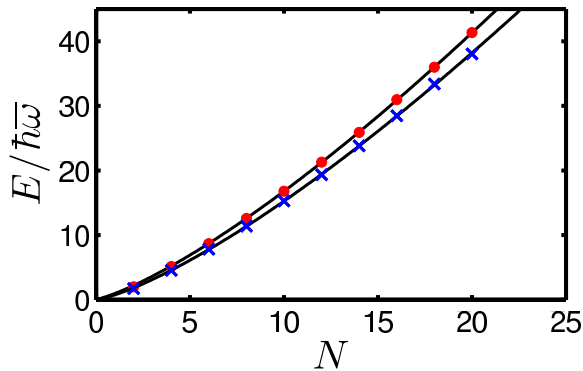}}
%\vspace*{-1.6in}
\caption{ (Color online) Circles and crosses show the FN-DMC
energies $E(N)$ in units of $\hbar \bar{\omega}$ as a function of
$N$ at unitarity for $\kappa=1$ ($\omega_1 = \omega_2$) and $\kappa
= 8$ ($\omega_2 = \kappa \omega_1$), respectively. Solid lines show
a fit of the FN-DMC energies to Eq.~(\ref{eq_tfexp}).} \label{fig4}
\end{figure}
for $\kappa=1$ and $8$, respectively, while
Table~\ref{energies} lists the FN-DMC energies.
In these calculations,
the range $R_0$ of the square well potential
used to describe the interaction between unlike fermions is set to
$R_0=0.01 a_{ho}^{(1)}$.
The energies for $N=4$ are calculated using $\psi_{T2}$; usage of
$\psi_{T1}$ leads to slightly higher energies.
As discussed already
in Sec.~\ref{fourbody}, the four-body FN-DMC energy
for equal masses
agrees well with the corresponding CG energy.
The energies for $N
\ge 6$ are calculated using the guiding function $\psi_{T1}$. For
example, usage of the guiding function $\psi_{T2}$ gives an
energy of $12.64(2) \hbar \bar{\omega}$ for $N=8$ and $\kappa=1$
(which is, taken the statistical errorbars into account, just
slightly higher than the energy calculated using $\psi_{T1}$; see
Table~\ref{energies}), and an energy of $43.2(1) \hbar \bar{\omega}$
for $N=20$ and $\kappa=1$ (which is notably higher than the energy
calculated using $\psi_{T1}$; see Table~\ref{energies}).

For $N>8$, our energies for equal masses and equal frequencies are
consistently lower than those reported in Ref.~\cite{chan07}. For
$N=20$, e.g., we find $E=41.35(8) \hbar \bar{\omega}$ while
Ref.~\cite{chan07} reports $43.2(4) \hbar \bar{\omega}$. We
speculate that this discrepancy can be traced back to the nodal
structure of the trial wave function employed, and possibly also to
the larger range of the two-body potential employed in
Ref.~\cite{chan07}. In agreement with Ref.~\cite{chan07}, we find
that the energies at unitarity show no shell structure.

Using the LDA,
which should be valid for sufficiently large $N$,
we relate the energy of the trapped
Fermi
system at unitarity
to the
universal parameter $\xi_{\kappa}$ and $E_{NI}$,
i.e.,
\begin{eqnarray}
E(N)=\sqrt{\xi_{\kappa}} E_{NI}.
\end{eqnarray}
% cut DB 04/23
The parameter $\xi_{\kappa}$ connects the energy per particle
$E_{hom}/N$ of the homogeneous system at unitarity and the energy
per particle $E_{FG}$ of the homogeneous non-interacting Fermi gas,
i.e., $E_{hom}/N = \xi_{\kappa} E_{FG}$. Here, we assumed that the
functional dependence of $E_{hom}/N$ on $E_{FG}$ is the same for
equal and unequal masses but that the universal parameter $\xi$
depends on $\kappa$. Applying the extended Thomas-Fermi (ETF) model
to $E_{NI}$~\cite{brack}, the energy of the trapped system at
unitarity becomes
\begin{eqnarray}
\label{eq_tfexp}
E(N) = \sqrt{\xi_{\kappa}}  \hbar \bar{\omega}
\frac{(3N)^{4/3}}{4}
\left(1+ c_{\kappa} \frac{(3N)^{-2/3}}{2} + \cdots \right) ,
\end{eqnarray}
where $c_{\kappa}=1$. The first term in Eq.~(\ref{eq_tfexp}) is
often referred to as Thomas-Fermi (TF) approximation. We also
attempted to fit our FN-DMC energies by functional forms different
from Eq.~(\ref{eq_tfexp}), which included higher-order correction
terms or terms with other powers of $N$; however, none of the
alternative functional forms considered improved the description of
our numerical results. Fitting our equal mass energies to
Eq.~(\ref{eq_tfexp}), treating $\xi_{\kappa}$ as a parameter, we
find $\xi_1=0.465$. Our $\xi_1$ extracted from the trapped system is
about 10\% larger than that determined for the bulk system, i.e.,
$\xi_{1}=0.42(1)$~\cite{astr04c,carl05}, suggesting that one has to
go to somewhat larger trapped systems to extrapolate the bulk
$\xi_1$ with high accuracy within the LDA. Although the $\xi_1$
obtained from the fit to the energies of the trapped system is
larger than the corresponding bulk value, it is worthwhile noting
that the simple functional form given in Eq.~(\ref{eq_tfexp})
provides an excellent description of the energies of the trapped
system.

For $\kappa=8$, we find that our energies are best described if we
treat $\xi_8$ and $c_8$ as fitting parameters, yielding $\xi_8=
0.417$ and $c_8=0.27$. The decrease of $c_8$ compared to the value
predicted by the ETF model is most likely related to the fact that
the densities of the unequal mass species do, in contrast to the LDA
treatment employed to derive Eq.~(\ref{eq_tfexp}), not fully overlap
at unitarity. Our calculations suggest that the bulk value for
$\xi_{\kappa}$ is somewhat smaller for $\kappa=8$ than for
$\kappa=1$. To investigate this further, we additionally consider
the energetics of systems with $\kappa=4$, 12, 16 and 20.
Figure~\ref{xi} shows the resulting $\xi_{\kappa}$ (main panel) and
$c_{\kappa}$ (inset)
\begin{figure}
\vspace*{.075in} \centerline{\epsfxsize=3in\epsfbox{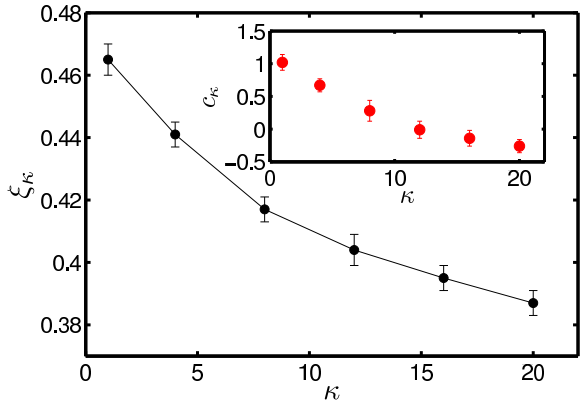}}
%\vspace*{-1.6in}
\caption{
(Color online)
$\xi_{\kappa}$ as a function of $\kappa$.
Inset: $c_{\kappa}$ as a function of $\kappa$.
The errorbars indicate the uncertainty of the fitting parameters;
this uncertainty does not include the statistical errors
of the FN-DMC energies.
}
\label{xi}
\end{figure}
extracted from our energies for $N=2-20$ fermions as a function of
$\kappa$. Both $\xi_{\kappa}$ and $c_{\kappa}$ vary smoothly,
decrease with increasing $\kappa$, and seem to approach a constant
for large $\kappa$. Furthermore, $c_{\kappa}$ changes sign from
positive to negative for $\kappa \approx 10$. We emphasize that
Eq.~(\ref{eq_tfexp}) provides a rather good description of the
energetics for all $\kappa$ thus empirically motivating the
non-constant coefficient $c_{\kappa}$ of the correction term.

The fact that $\xi_{\kappa}$ decreases with increasing
$\kappa$ is in agreement
with recent FN-DMC calculations for the homogeneous system~\cite{astr07}.
However, this decrease is more pronounced for the trapped
system than for the  homogeneous system.
Standard BCS
mean-field theory, in contrast, predicts that the parameter $\xi$, which
determines many properties of dilute homogeneous
two-component Fermi gases at unitarity~\cite{ho04},
is independent of $\kappa$~\cite{wu06}.

We now comment further on the choice of the guiding function used to
obtain the energies for $N=2-20$ that enter into our determination
of $\xi_{\kappa}$ and $c_{\kappa}$. As mentioned earlier, trimer
states with negative energy exist for $\kappa=16$ and $20$. If we
use the guiding function $\psi_{T2}$ to model these systems,
many-body configurations that contain three particles in close
proximity are being sampled, giving rise to negative energies at
unitarity. However, if we use the guiding function $\psi_{T1}$,
i.e., if we construct the nodal surface by pairing spin-up and
spin-down particles, many-body configurations that contain
tightly-bound trimer states are not being sampled. Thus, the guiding
function $\psi_{T1}$ allows for a numerically stable
characterization of a state with positive energy that has the same
symmetry as $\psi_{T1}$. We note that the decrease of $\xi_{\kappa}$
with increasing $\kappa$ is already present in the energies for the
two-body system for which we can determine the energetics
essentially exactly. This provides some evidence that the behaviors
discussed in this section for unequal masses are not an artefact of
our choice of guiding function.

The difference between the guiding functions $\psi_{T1}$ and
$\psi_{T2}$ can also be understood from a different point of view.
In the limit of vanishing confinement, the oscillator states used to
construct $\psi_T$ approach free-particle states. In this case, the
nodal surface of the guiding function $\psi_{T1}$ is compatible with
a superfluid state, and the guiding function $\psi_{T2}$ with a
normal state~\cite{carl03,chan04,astr04c,carl05}. Using this
analogy, our results suggest that even fairly small trapped Fermi
systems are better described by a ``superfluid wave function'' than
by a ``normal wave function''. As $N$ increases, the difference
between the energies obtained for the superfluid and normal wave
functions increases, presumably approaching the bulk values in the
large $N$ limit ($E_{hom}/N=0.42(1) E_{FG}$~\cite{carl05,astr04c}
for the superfluid state and $E_{hom}/N = 0.54 E_{FG}$ for the
normal state~\cite{carl03,carl05}).

\section{Conclusion}
\label{conclusion} This paper characterizes the BEC-BCS crossover
physics of trapped two-component Fermi gases with varying mass
ratio. Our results are obtained by solving the stationary many-body
Schr\"odinger equation for short-range model potential by two
complementary approaches. For the four-particle system, an
essentially exact basis set expansion type technique, a CG approach,
is used to determine the complete $L^P=0^+$ spectrum. For up to
$N=20$ particles, the FN-DMC approach is used to determine upper
bounds for energy of the BEC-BCS crossover branch. Treating the
four-body system is challenging, and interesting in its own right:
The four-body system is the smallest non-trivial system exhibiting
BEC-BCS crossover-like physics. Furthermore, the lessons learned
from the four-body system aid the study of larger systems. Solving
the Schr\"odinger equation for more than a few fermions by first
principle methods is, despite the increasing available computer
power, still a challenging task. In fact, it may be argued that
Monte Carlo methods are the only methods suitable. Unfortunately
however, assessing the accuracy of the assumptions going into Monte
Carlo calculations, such as the nodal surface employed in the FN-DMC
approach, remains a challenge. Our calculations, which use the CG
and FN-DMC approaches in parallel, benchmark the strengths and
limitations of the nodal surface employed in the FN-DMC
calculations.

From our four-body calculations in the deep BEC regime, we determine
the scattering length $a_{dd}$ and effective range $r_{dd}$ of the
dimer-dimer system for two purely attractive short-range two-body
potentials as a function of the mass ratio $\kappa$. For up to
$\kappa \approx 13.6$, our dimer-dimer scattering lengths $a_{dd}$
agree well with the values calculated by Petrov {\em{et
al.}}~\cite{petr05}. Our four-body calculations extend beyond this
mass ratio $\kappa$ and suggest that the energetically lowest-lying
dimer-dimer state varies smoothly as a function of $\kappa$. We find
that the energies of the dimer-dimer states for large $\kappa$
depend, just as for small $\kappa$, at most weakly on the details of
the two-body potential; we take this as numerical evidence that at
least some properties of the dimer-dimer states are universal even
if $\kappa$ exceeds the value of 13.6. However, other properties of
systems with large $\kappa$ such as the system's stability,
encapsulated in the three-body recombination rate, are presumably
controlled by the details of the short-range potential. These
non-universal properties should be investigated in the future. Also,
future studies will have to investigate whether the comparatively
large value of $r_{dd}$ can be measured indirectly by, e.g., a
careful analysis of the density profile in the BEC regime.

We also present calculations in the strongly-interacting unitarity
regime for different mass ratios. Our calculations for $N=2-20$
fermions show no shell structure. Application of the LDA to the
trapped system implies that the universal parameter $\kappa$ depends
weakly on the mass ratio. Our energies at unitarity for various mass
ratios may aid in developing and refining numerically less demanding
treatments of two-component Fermi gases. Recently, e.g.,
Bulgac~\cite{bulgac07} proposed a density functional theory
applicable to trapped equal-mass two-component Fermi gases at
unitarity. Our results presented here provide much needed benchmarks
for such theories. Our unequal mass studies present the first first
principle treatment of such systems under confinement. Our analysis
provides a first step towards a deeper understanding of these
systems, but much room for further investigations, including the
investigation of connections to mean-field
treatments~\cite{wu06,cald05,lin06,iski06,iski06a,pari06,he06},
remains.

We gratefully acknowledge discussions with S. Giorgini and S.
Rittenhouse, and support by the NSF.

%under grants
%No. PHY-0331529 and xxx.


\begin{thebibliography}{10}

\bibitem{rega03a}
C.~A. Regal, C. Ticknor, J.~L. Bohn, and D.~S. Jin, Nature {\bf
424},  47
  (2003).

\bibitem{zwie03}
M.~W. Zwierlein, C.~A. Stan, C.~H. Schunck, S.~M.~F. Raupach, S.
Gupta, Z.
  Hadzibabic, and W. Ketterle, Phys. Rev. Lett. {\bf 91},  250401  (2003).

\bibitem{stre03}
K.~E. Strecker, G.~B. Partridge, and R.~G. Hulet, Phys. Rev. Lett.
{\bf 91},
  080406  (2003).

\bibitem{joch03}
S. Jochim, M. Bartenstein, A. Altmeyer, G. Hendl, S. Riedl, C. Chin,
J. {Hecker
  Denschlag}, and R. Grimm, Science {\bf 302},  2101  (2003).


\bibitem{footnote}
{In the deep BEC regime, i.e., for small negative $s$-wave scattering
length $a_s$, the energies of the BEC branches depend in general on
the  range $R_0$ of the two-body potential. This dependence can,
in principle, be eliminated by taking the limit $R_0 \rightarrow 0$.
Throughout this work, we do not evaluate the $R_0 \rightarrow 0$
limit for numerical reasons, and instead subtract the dimer binding
energy [see Eq.~(5)]. Consequently, we refer to states of the BEC
branch as universal states if their energy, with the dimer binding
energy subtracted, is independent of the range of the two-body
potential.}

\bibitem{efim70}
V.~N. Efimov, Yad. Fiz. {\bf 12},  1080  (1970 [Sov. J. Nucl. Phys.
12, 589
  (1971)]).

\bibitem{efim73}
V.~N. Efimov, Nucl. Phys. A {\bf 210},  157  (1973).

\bibitem{petr05}
D.~S. Petrov, C. Salomon, and G.~V. Shlyapnikov, J. Phys. B {\bf
38},  S645
  (2005).

\bibitem{bulgac07}
A. Bulgac, {cond-mat/0703526}  (2007).

\bibitem{singer1960uge}
K. Singer, Proc. R. Soc. London, Ser. A {\bf 258},  412  (1960).

\bibitem{suzuki1998sva}
Y. Suzuki and K. Varga, {\em {Stochastic Variational Approach to
  Quantum-Mechanical Few-Body Problems}} (Springer-Verlag, Berlin, 1998).

\bibitem{stec07}
J. {von Stecher} and C.~H. Greene, {cond-mat/0701044}  (2007).

\bibitem{reyn82}
P.~J. Reynolds, D.~M. Ceperley, B.~J. Alder, and {W. A. Lester,
Jr.}, J. Chem.
  Phys. {\bf 77},  5593  (1982).

\bibitem{hamm94}
B.~L. Hammond, W.~A. {Lester, Jr.}, and P.~J. Reynolds, {\em Monte
Carlo
  Methods in Ab Initio Quantum Chemistry} (World Scientific, Singapore,
   1994).

\bibitem{astr04c}
G.~E. Astrakharchik, J. Boronat, J.~D. Casulleras, and S. Giorgini,
Phys. Rev.
  Lett. {\bf 93},  200404  (2004).

\bibitem{footnote4}
{The terms spin-up and spin-down are used figuratively. For equal
masses, these
  terms refer to two atoms of the same isotope trapped in two different
  internal (hyperfine) states. For unequal masses, these terms refer to two
  different fermionic atoms, each trapped in a suitable internal (hyperfine)
  state.}

\bibitem{legg80}
A.~J. Leggett, {\em Modern Trends in the Theory of Condensed Matter,
ed. by A.
  Pekalski and R. Przystawa} (Springer Verlag, Berlin, 1980).

\bibitem{nozi85}
P. Nozi\`eres and S. Schmitt-Rink, J. Low Temp. Phys. {\bf 95},  195
(1985).

\bibitem{enge97}
J.~R. Engelbrecht, M. Randeria, and C.~A.~R. {S\'a de Melo}, Phys.
Rev. B {\bf
  55},  15153  (1997).

\bibitem{astr07}
G.~E. Astrakharchik, D. Blume, and S. Giorgini, {unpublished}
(2006/2007).

\bibitem{huan57}
K. Huang and C.~N. Yang, Phys. Rev. {\bf 105},  767  (1957).

\bibitem{blum02}
D. Blume and C.~H. Greene, Phys. Rev. A {\bf 65},  043613  (2002).

\bibitem{bold02}
E.~L. Bolda, E. Tiesinga, and P.~S. Julienne, Phys. Rev. A {\bf 66},
013403
  (2002).

\bibitem{busc98}
T. Busch, B.-G. Englert, K. Rz\c{a}\.zewski, and M. Wilkens,
Foundations of
  Phys. {\bf 28},  549  (1998).

\bibitem{fermi36}
E. Fermi, Ric. Sci. {\bf 7},  13  (1936).

\bibitem{vonstech07}
J. von Stecher and C.~H. Greene, Phys. Rev. A {\bf 75},  022716
(2007).

\bibitem{chan07}
{S.~Y. Chang and G.~F. Bertsch}, {physics/0703190}  (2007).

\bibitem{brack}
M. Brack and R.~K. Bhaduri, {\em Semiclassical Physics}
(Addison-Wesley,
  Reading, MA, 1997).

\bibitem{carl05}
J. Carlson and S. Reddy, Phys. Rev. Lett. {\bf 95},  060401  (2005).

\bibitem{ho04}
T.-L. Ho, Phys. Rev. Lett. {\bf 92},  090402  (2004).

\bibitem{wu06}
S.-T. Wu, C.-H. Pao, and S.-K. Yip, Phys. Rev. B {\bf 74},  224504
(2006).

\bibitem{carl03}
J. Carlson, S.~Y. Chang, V.~R. Pandharipande, and K.~E. Schmidt,
Phys. Rev.
  Lett. {\bf 91},  050401  (2003).

\bibitem{chan04}
S.~Y. Chang, V.~R. Pandharipande, J. Carlson, and K.~E. Schmidt,
Phys. Rev. A
  {\bf 70},  043602  (2004).

\bibitem{cald05}
H. Caldas, C.~W. Morais, and A.~L. Mota, Phys. Rev. D {\bf 72},
045008
  (2005).

\bibitem{lin06}
G.-D. Lin, W. Yi, and L.-M. Duan, Phys. Rev. A {\bf 74},  031604
(2006).

\bibitem{iski06}
M. Iskin and C.~A. R.~S\'a de~Melo, cond-mat/0604184  (2006).

\bibitem{iski06a}
M. Iskin and C.~A. R.~S\'a de~Melo, cond-mat/0606624  (2006).

\bibitem{pari06}
M.~M. Parish, F.~M. Marchetti, A. Lamacraft, and B.~D. Simons, Phys.
Rev. Lett.
  {\bf 98},  160402  (2007).

\bibitem{he06}
L. He, M. Jin, and P. Zhuang, Phys. Rev. B {\bf 73},  220504(R)
(2006).

\end{thebibliography}
\end{document}